\documentclass[twocolumn,showpacs,preprintnumbers,amsmath,amssymb]{revtex4}

\usepackage{graphicx}
\usepackage{dcolumn}
\usepackage{bm}
\newcommand{\ableit}[2]{\frac{\partial #1}{\partial #2}}

\renewcommand{\)} {\right )}

\begin{document}

\title{Quantum signatures in laser-driven relativistic multiple-scattering}

\author{Guido R. Mocken}%
\email{mocken@physik.uni-freiburg.de}
\author{Christoph H. Keitel}%
\email{keitel@uni-freiburg.de}\homepage{http://tqd1.physik.uni-freiburg.de/~chk/a11/index_de.html
}

\affiliation{%
Theoretische Quantendynamik, Physikalisches Institut, Universit\"at Freiburg,\\
Hermann-Herder-Stra{\ss}e 3, D-79104 Freiburg, Germany
}%
\date{\today}

\begin{abstract}
The dynamics of an electronic Dirac wave packet evolving under the influence of an ultra-intense laser pulse and an ensemble of 
highly charged ions is investigated numerically. Special emphasis is placed on the evolution of quantum signatures from  
single to multiple scattering events. We quantify the occurrence of quantum relativistic interference fringes in various situations 
and stress their significance in multiple-particle systems, even in the relativistic range of laser-matter interaction. 

\end{abstract}

\maketitle

The interplay of the strongest forces in atomic physics via ultra intense laser pulses and highly charged ions is governed 
rather well by quantum relativistic Dirac dynamics \cite{reviewions,reviewfields,Maquet:2002}. 
On one hand, for single particles quantum effects such as tunneling, 
spin effects and quantum interferences have shown to be rather crucial even in the regime of 
ultra-short and highly relativistic dynamics \cite{Rathe:1997,Panek:2002}. On the other hand, for many particle systems, laser-induced plasma 
physics was shown to be remarkably well described by classical relativistic dynamics \cite{plasmaex,plasmathe}. With the intermediate regime 
from few particle to cluster physics attracting increasing interest \cite{cluster}, the question arises for the role of quantum effects in laser-induced 
relativistic dynamics.

In this letter we investigate the quantum relativistic dynamics of laser-driven multiple-scatterings of an electron being 
represented by a Volkov wave packet at an ensemble of highly charged ions. With an numerical accuracy, which allows 
for transitions even to the Dirac sea with negative energies, we quantify the interference fringes at each scattering event 
and the mutual interplay among those events. Clear quantum behaviour in the Dirac wave packet is identified 
in the highly relativistic regime after multiple scattering.

The system of interest consists of an electron which is driven by an intense laser pulse with 
time $t$ and space $\vec r$ dependent vector potential ${\vec A}(t, \vec r \,)$ and scattered multiply 
at an ensemble of ions with scalar potential  $A_0(\vec r \,)$.
The electronic wave packet dynamics in such an environment is characterised by the Dirac  
spinor $\psi({t,\vec r} \,)$ and is governed by the Dirac equation reading in atomic units 
as throughout the article: 
	\begin{eqnarray}
		 { i} \hbar \ableit{\psi}{t}  = \left[ {\hbar c \over { i}} \alpha^j  \ableit{}{r^j} + \beta mc^2 +q\(A_0 - \alpha^j A_j \)\right] \psi 
		 \label{eq:time_evolver}	\end{eqnarray}
with electron charge $q = -1 $a.u., electron mass $ m= 1$ a.u.   and $\alpha^j$ ($j\in \{1,2,3\}$) and $\beta$ being the  
Dirac matrices \cite{Bjorken_Drell}. 
The three components of $\vec r$ and $\vec A$ are $r^j$ and $A^j$, respectively, and $c=137.036$ a.u. the speed of light.

\begin{figure*}
	\includegraphics[scale=0.8215, clip]{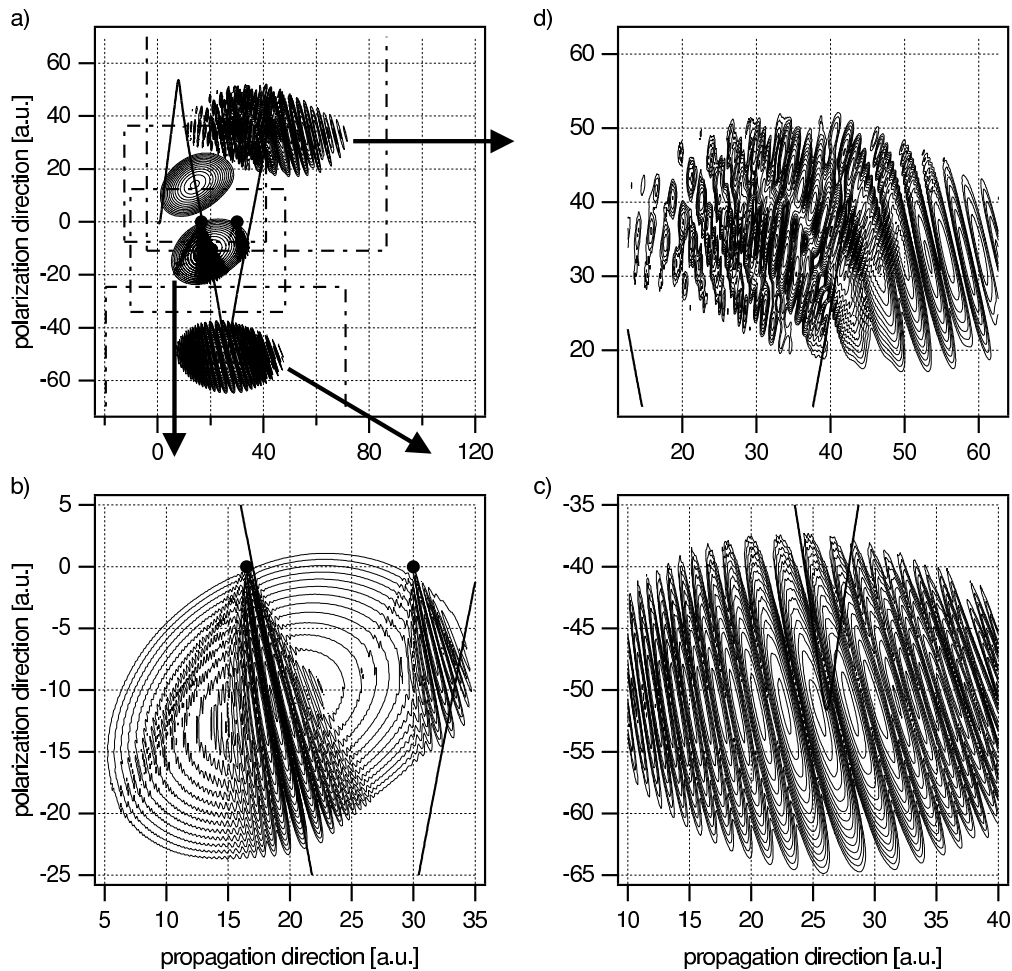}%
	 \caption{
	 a) Overview: Contour plots at $t= 7.645, 8.195, 9.495, 12.095$ a.u. corresponding to times before the first scattering, shortly after it, 
	 at the lower point of return, and after the second scattering, close to the upper point of return. The four dashed rectangles (some only partly 
	 visible) mark the dynamical grid boundaries for each of these situations. The thick dots indicate the positions of two $\text{Sn}^{50+}$ ions, 
	 which are located right on the horizontal axis at positions $16.5$ and $30.0$ a.u. The solid line depicts the trajectory of the expectation value 
	 of the particle's spatial coordinate. We omit any enlarged view of the initially Gaussian packet at $t= 7.645$, but provide them for the other 
	 three cases as follows: 
	 b) Enlarged view at $t= 8.195$ a.u. showing single scattering fringes shortly after the scattering event;
	 c) Enlarged view at $t= 9.495$ a.u. that depicts the same single scattering fringes somewhat later in order to illustrate the growth of the 
	 distance between any two fringes (compare with b));
	 d) Enlarged view at $t= 12.095$ a.u. which displays interference from two separate scattering events with crossed fringes on the left side and 
	 the unperturbed structure from single scattering at the first ion only on the right side.
	 In all cases contour lines are shown for $|\psi|^2$ with $\log|\psi|^2\ge {-4}$ and line spacings marking steps of  $0.15$.
	 } 
	\label{fig:Igor3in1}
\end{figure*}

Our numerical analysis takes advantage of splitting the linear Dirac Hamiltonian into a position-dependent and a derivative-dependent part.
We then make use of the so-called ``split-operator'' technique \cite{Fleck:1976}, in which we propagate the wave packet successively 
by the position- and  derivative-dependent parts and employ fast-Fourier-transformations, such that all operations are plain multiplications.
With time step $\Delta t$, the numerical error introduced this way is of the order  $\Delta t^3$ \cite{Fleck:1976}.
For $ \Delta t \approx 2\times 10^{-5}$ a.u.$ < \frac{\hbar}{2mc^2} $, transitions between positive- and negative-energy states are resolved, 
and this way we obtained convergence of our split-step propagation of $\psi$ even at large $t\gtrsim10$ a.u.. 
Further the spacing of the grid in position space needs to be suitable to resolve the maximal momenta employed. In spite of large relativistic velocities,
this is not problematic, because it is the canonical rather than the kinetic momentum that has to be represented on 
the grid, with $\vec p_{can} -\frac{q}{c}\vec A = \vec p_{kin}$. In the case of a high velocity of the particle being exclusively due 
to an intense laser field, $\vec p_{can}$ is even zero in polarization direction. 
It is non-zero in propagation direction, but its magnitude is small even for intense fields. Once scatterings with nuclei have occurred,
however high canonical momenta appear, which, for the parameters used 
here, can be represented successfully on a grid with a spacing $\Delta x_i=0.118$ a.u., corresponding to a maximum momentum of $26.6$ a.u.. 
The so-called fermion doubling problem, which occurs at momenta close to the highest grid momenta, is consequently also avoided \cite{Mueller:1998}.
For the sake of reducing computing power, we introduced two advantageous techniques. Firstly, in position space, the calculation is restricted to 
the area centered around the rapidly moving wave packet, involving a ``moving-grid" approach.
Secondly, the grid size, too, is dynamically adapted in time: While a freely evolving wave packet spreads with time, a multiply scattered one does considerably 
more. As our simulation has to cover a substantial part of the whole laser pulse, including times where the packet is still quite small, it is possible 
to save considerable CPU time, noting that the time consuming two-dimensional fast-Fourier-transformations scale as $O(N^2 \log N)$, where $N \times N$ equals 
the grid size. This ``growing-grid" approach is also our pragmatic solution to the well-known boundary problem \cite{Alonso:1997, Maquet:2002} 
in Dirac calculations, at least to the point where damping functions and absorbing boundaries become unavoidable.
Finally, the whole code is written to take advantage of multiple CPUs.

In a series of contour plots, we present the time evolution of an initially Gaussian shaped wave packet under the influence of a strong laser pulse, 
which is subsequently scattered at several highly charged ions.
We use a four cycle laser pulse with amplitude $E_0=50$ a.u. ($I=8.75\times 10^{19}\frac{W}{cm^2}$) and frequency $\omega=1$ a.u., 
which features a 1.5 cycles $sin^2$ turn-on and turn-off, and one cycle of constant intensity in between. 
As amplitude and frequency suggest, we are clearly in the fully relativistic regime.
The ions are modeled by static softcore potentials $Ze^2 / \sqrt{((\vec r-\vec r_\text{Ion})^2+a)}$, with a ``Coulomb-like'' 
small softcore parameter $a=0.01$ being just large enough to avoid numerical instabilities at the ions' origins $\vec r_{\rm Ion}$. 
We chose their charge as a high multiple ($Z=50$) of the elementary charge $e$ in order to acquire comparable field strengths for laser 
and ions (at 1 a.u. distance from the ionic center).

In fig. \ref{fig:Igor3in1}, the top left graph illustrates an overview of the successive quantum relativistic scattering scenario of 
an electron wave packet at two highly charged ions. The initial Gaussian wave packet (positive energy and spin up only) 
is centered around the origin and its evolution is depicted by various snap shots along its center of mass motion (solid line) in 
the laser pulse. 
At first, after a short motion in the negative polarization and positive propagation directions during the first half cycle of the 
turn-on phase, the particle is visibly accelerated in the polarization direction, reaching the first upper turning point 
after the first whole cycle is completed. 
Further on, continuing with a clear Lorentz-force induced drift in the laser propagation direction, the electron wave packet 
is accelerated in the negative polarization direction to face its first encounter with an ionic core potential. 
The motion that we observe during the first unperturbed $1\frac{1}{4}$ cycles could be modeled rather accurately and easily 
with a classical Monte Carlo ensemble and is completely in agreement with known free (without nucleus) quantum wave packet results 
in \cite{Roman:2001}. This includes the Lorentz-contraction of the wave packet along the direction of present velocity and its apparent rotation, 
i.e. precisely shearing, because of the phase differences sensed by spatially separated parts of the wave packet.

At the first encounter of the electron wave packet with the nuclei at position $(16.5, 0)$, the laser electric field is rather small 
at this particular phase.  Therefore relativistic Coulomb scattering dominates, involving the interference of the incoming wave with the scattered 
wave. The corresponding fringe structure which features a distinct maximum in forward direction followed by a series of side maxima, can be viewed 
in  detail in fig. \ref{fig:Igor3in1}b. Further fringes from a scattering at the second ion at $(30, 0)$ are also visible. 
With increasing time the wave packet evolves in the negative polarization direction towards the second lower turning point, which is 
reached after 1.5 cycles. The separations among the fringes have grown substantially which themselves turn continuously more parallel, 
with an orientation reflecting the direction of motion at the time when the scattering occurred (fig. \ref{fig:Igor3in1}c).

Finally, the wave packet, which at this stage is split in various sub-wave packets, continues in positive polarization direction. 
The fringes maintain their orientation when the second scattering at the second ion occurs, and are therefore joined by a set of newly created fringes 
with a different orientation. When two cycles are completed, the electron has reached its second upper turning point. Fig. \ref{fig:Igor3in1}d is 
a snapshot taken a little while before that, and one can clearly see the pattern that is generated by two sets of differently oriented interference 
fringes.
>From then on, without further scatterings, the whole structure essentially remains, apart from further changes imposed by the
laser field such as spreading and shearing with an essentially classical center-of-mass motion.

\begin{figure*}
	\includegraphics[scale=0.8, clip]{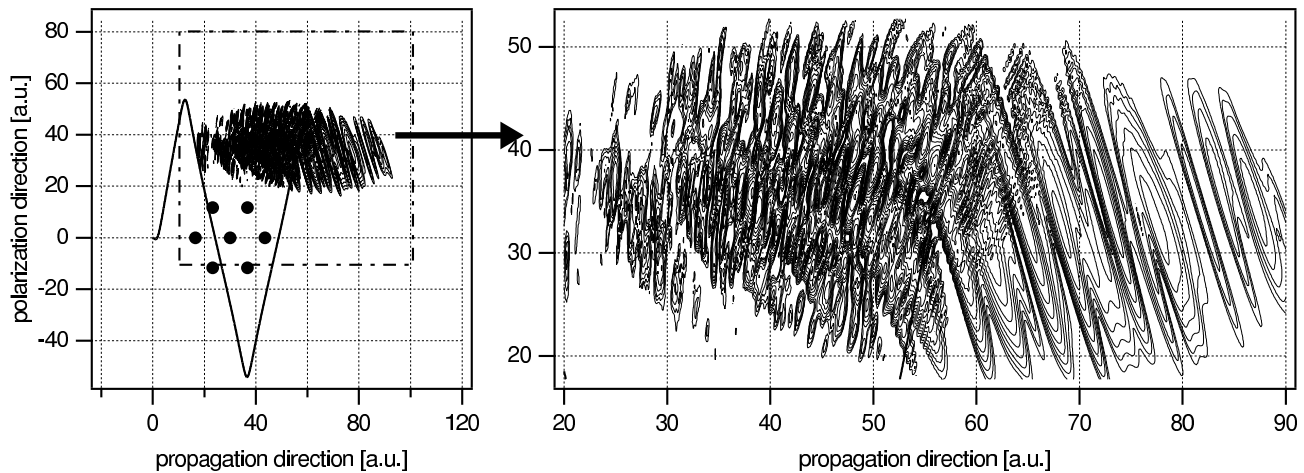}%
	 \caption{a) Overview: Contour plots at time $t= 12.270$ a.u. after two scatterings at an ensemble of six $\text{ Sn}^{50+}$ 
	 ions (thick dots) centered symmetrically around a further one at position  $(0, 30)$ a.u.. 
	 The dashed rectangle marks the grid boundary and the solid line depicts the trajectory of the expectation value of the electron's 
	 spatial coordinate.	 b) This enlarged view  of the wave packet in a) illustrates how the two scattering events at seven ions 
	 modify the regularity in the interference pattern. 
          Contour lines are shown for $|\psi|^2$ with $\log|\psi|^2\ge {-4}$ and line spacings marking steps of  $0.15$.}
	\label{fig:seven}
\end{figure*}

In figure \ref{fig:seven}, we show in a similar way the effect on the same initially Gaussian wave packet after it has passed a collection of seven 
ions two times. A classical estimation confirms that such a collection of heavy highly charged ions moves less than 0.04 a.u. due to Coulomb repulsion 
and the laser field during maximally employed interaction time of $\Delta t \approx 12 $ a.u. and may thus be assumed as resting.
We note that the clear interference structures in single and double scattering are now less apparent though still visible 
when one moves to complex structures. 

A simple analytical model is finally introduced to qualitatively confirm our numerical result.
Adopting existing text-book \cite{Strange} theory for three-dimensional Dirac scattering of an electron $\psi$
at the time-independent potential of an ion at $\vec r_{\text {Ion}}=\vec 0$, we obtain

\begin{equation}
\label{Green_0_definition}
	\psi(\vec r\,) = \phi(\vec r\,) -  \int d^3r' G_0(\vec r, \vec r\,', E)V(\vec r\,')\psi(\vec r\,') 
\end{equation}
with unperturbed Dirac wave  $\phi(\vec r\,) = w^\rho(\vec p\,) e^{\frac{i}{\hbar}\vec p \cdot \vec r}$ ($\rho\in\{1,2\}$),
corresponding eigenvalue $E$ and $w^\rho(\vec p\,)$ being the free-electron spinor amplitude \cite{Bjorken_Drell}. 
 On the right hand side of eq. (\ref{Green_0_definition}), we replace $\psi$ by $\phi$ (first Born approximation) and  
 insert the relativistic free-particle Green's function at energy $E$ \cite{Strange}
\begin{equation}
\label{Green_0_free}
	G_0(\vec r, \vec r\,', E) = \frac{1}{\hbar c}\left[c \vec\alpha \cdot \vec {\underline{p}} + \beta mc^2 + E\right] \frac{e^{\frac{i}{\hbar}pR}}{4\pi R}
,\end{equation}
where $R=|\vec r -\vec r\,'|$, $p=\sqrt{\frac{E^2}{c^2}-(mc)^2}$, $\vec{\underline{p}} = -i\hbar\vec\nabla$, $\vec p$  the initial momentum, 
$\vec p\,' = p \frac{\vec r}{r}$ the final momentum and $p=\hbar k=|\vec p\,|=|\vec p\,'|$ its magnitude. 
For the case of interest $r\gg r'$, neglecting contributions of order $\frac{1}{r^2}$ and higher, and assuming a 
short-range potential, one finally obtains the outgoing electronic wavefunction

\begin{eqnarray}
\label{psi_asymptotisch}
	\psi(\vec r\,) \mkern-10mu&=& \mkern-7mu w^\rho(\vec p\,) e^{\frac{i}{\hbar}\vec p 
	\cdot \vec r} - \frac{1}{4\pi (\hbar c)^2}\biggl[\vec \alpha \cdot \vec p\,'  + \beta mc^2 +  \nonumber\\
	& & \mkern-7mu E(\vec p\,'^2)\biggr]\frac{e^{\frac{i}{\hbar} p|\vec r\,|}}{|\vec r\,|} \mkern-7mu\int \mkern-7mu d^3r' 
	V(\vec r\,') w^\rho(\vec p\,) e^{\frac{i}{\hbar}(\vec p-\vec p\,') \cdot \vec r\,'}\mkern-7mu.
\end{eqnarray}
We are interested in the maxima of $|\psi|^2=\psi^{\dagger} \psi$, or more exactly in the angles $\vartheta_{n}$ that point towards the scattering fringes. Using 
	$V(\vec r\,') = - V_0 \delta(\vec r\,')$ \cite{becker} with $V_0 >0$
as the simplest potential,
we finally obtain, up to an additive function $f(r)$ and a constant pre-factor, the $\vartheta$-dependant part of $|\psi|^2$ as
\begin{equation}
	 |\psi|^2 \propto ( (\gamma^2-1) \cos\vartheta  + 1 + \gamma^2) \cos\left( k r -  k r \cos\vartheta \right) + f(r).
\end{equation}
In the nonrelativistic case  $\gamma = \frac{E}{mc^2} \approx 1$, the maxima of the above expression can be simplified  further to 
read 
\begin{equation}
	\vartheta_{n}= \pm \arccos\left(1-\frac{n\pi}{k r }\right), \quad n \in \mathbb N.
\end{equation}
Then to adapt the dynamics in the laser fields, one may choose for a fixed initial momentum $\hbar k$ a distance $r$ where, in the \emph{absence} of a 
laser field, the scattering fringes would be observed and calculate the corresponding angles $\vartheta_{n}$. Then with the laser field and, 
using a classical formula \cite{Salamin:1996} and now neglecting the ionic potentials, one may propagate over a period $t=\frac{m r}{\hbar k}$ a suitably 
chosen ensemble of classical particles that initially starts at the position of the scattering 
center with initial momenta of magnitude $\hbar k$ in the direction of the scattering angles $\vartheta_{n}$.
This simple model qualitatively confirms our numerical results while it fails to predict the final positions and separations 
of the fringes by better than a factor of two. In addition to the stressed approximations in the analytical approach, the 
mimicking of the quantum wave packet in the transition regime from scattering to free dynamics in the laser field is too 
delicate to compete seriously in accuracy with the up-initio quantum relativistic approach.

Concluding, relativistic quantum dynamics was investigated for a multiple-particle system with clear interference fringes 
being identified and quantified. While quantum signatures in many-particle systems are likely to be washed out, our examples 
show that there is an intermediate regime in the number of involved particles with clear quantum effects for relativistic dynamics. 

Financial support by the German Science Foundation (Nachwuchsgruppe within SFB276) is gratefully acknowledged.

\hyphenation{Post-Script Sprin-ger}

\end{document}